\documentstyle[11pt]{article}
\newcommand{\ket}[1]{\left | #1 \right \rangle}
\newcommand{\bra}[1]{\left \langle #1 \right |}
\newcommand{\amp}[2]{\left \langle #1 | #2 \right \rangle}
\newcommand{\proj}[1]{\ket{#1} \bra{#1}}

\newcommand{\reals}{\mbox{I$\!$R}}
\begin{document}
\begin{center} {\bf SEARCHING IN GROVER'S ALGORITHM}\\ Richard
Jozsa\\School of Mathematics and Statistics\\University of
Plymouth\\Plymouth, Devon PL4 8AA, England.\\Email:
rjozsa@plymouth.ac.uk
\end{center}

\noindent
{\small {\bf ABSTRACT}

\noindent
Grover's algorithm is usually described in terms of the iteration
of a compound operator of the form $Q=-HI_{0}HI_{x_0}$. Although it
is quite straightforward to verify the algebra of the iteration,
this gives little insight into why the algorithm works. What is the
significance of the compound structure of $Q$? Why is there a minus
sign? Later it was discovered that $H$ could be replaced by
essentially any unitary $U$. What is the freedom involved here? We
give a description of Grover's algorithm which provides some
clarification of these questions.}

\noindent
{\bf INTRODUCTION}

Grover's quantum searching algorithm is usually described
\cite{GRO96,GRO97,GRO97a,BOY96} in terms of the iteration of a
compound operator $Q$ of the form
\begin{equation} \label{qh} Q = -HI_0 H I_{x_0} \end{equation}
on a starting state $\ket{\psi_0} = H\ket{0}$. Here $H$ is the
Walsh-Hadamard transform and $I_0 ,I_{x_0}$ are suitable inversion
operators (c.f. below).\footnote{ This $Q$ is the one used in
\cite{BOY96}. Grover \cite{GRO96,GRO97} uses instead $Q^{GR}=
-I_0 HI_{x_0} H$ iterated on $\ket{\psi_0}= \ket{0}$ which is clearly
an equivalent process.} Later it was discovered
\cite{BOY96,GRO98,BRA98} that $H$ may be replaced by essentially
{\it any} unitary operation $U$ and using
\begin{equation} \label{qu}
Q = -UI_0 U^{-1} I_{x_0} \hspace{1cm} \ket{\psi_0} = U\ket{0}
\end{equation}
the searching algorithm still works just as well as before (with at
most a constant slowdown in the number of iterations). At first
sight this appeared remarkable since $H$ is known to be singularly
significant for other quantum algorithms \cite{DEU92,SIM94,JOZ98b}.
The efficacy of these other algorithms could be understood in terms
of the fast Fourier transform construction \cite{EKE98} but
Grover's algorithm appears to rest on different principles.
Although it is quite straightforward to work through the algebra of
the algorithm \cite{GRO96,BOY96}, this provides little insight into
why it works! The operator $-HI_0 H$ was originally called a
``diffusion'' operator \cite{GRO96} and later interpreted as
``inversion in the average'' \cite{GRO97} but neither of these
appears to provide much heuristic insight (especially in the
context of the more general eq. (\ref{qu})). What is the
significance of the particular compound structure of $Q$ for a {\it
searching} problem? What is the significance of the minus sign in
$Q$? Why can $H$ be replaced by an arbitrary $U$
-- what is the freedom involved here? The purpose of this note is
to give a different description of Grover's algorithm which
provides some clarification of these issues. We will show that the
algorithm may be seen to be a consequence of the following
elementary theorem of 2-dimensional real Euclidean geometry:\\[2mm]
{\bf Theorem 1:} Let $M1$ and $M2$ be two mirror lines in the
Euclidean plane $\reals^2$ intersecting at a point $O$ and let
$\alpha$ be the angle in the plane from $M1$ to $M2$. Then the
operation of reflection in $M1$ followed by reflection in $M2$ is
just rotation by angle $2\alpha$ about the point $O$.\\
\begin{picture}(12,7)(0,0)
\put(2.6,0.6){\line(1,1){5.3}}
\put(2.0,1.4){\line(3,1){6.9}}
\put(5.0,2.65){$\alpha $}
\put(4.05,1.7){$O$}
\put(9.3,3.65){$ M1 $}
\put(8.2,5.9){$ M2 $}
\end{picture}
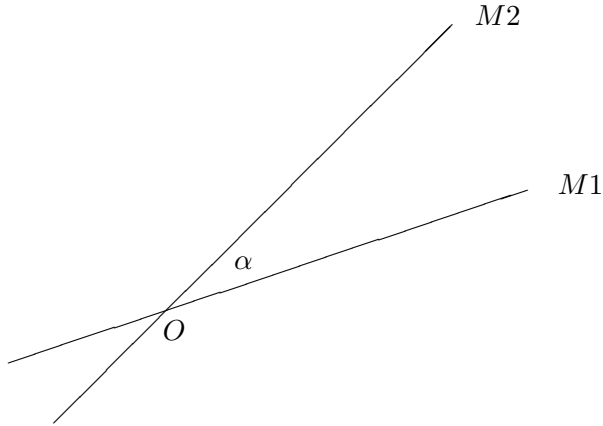
\\{\small Figure 1. Reflection in $M1$ followed by reflection in $M2$
is equivalent to rotation about $O$ through angle
$2\alpha$}.\\[2mm]

\noindent
{\bf THE SEARCH PROBLEM}

The search problem is often phrased in terms of an exponentially
large unstructured database with $N=2^n$ records, of which one is
specially marked. The problem is to locate the special record.
Elementary probability theory shows that classically if we examine
$k$ records then we have probability $k/N$ of finding the special
one so we need $O(N)$ such trials to find it with any constant
(independent of $N$) level of probability. Grover's quantum
algorithm achieves this result with only $O(\sqrt{N})$ steps (or
more precisely $O(\sqrt{N})$ iterations of $Q$ but $O(\sqrt{N} \log
N )$ steps, the $\log N$ term coming from the implementation of
$H$.) It may be shown \cite{BEN97} that the square root speedup of
Grover's algorithm is optimal within the context of quantum
computation.

The search problem may be more accurately phrased in terms of an
oracle problem, which we adopt here. In the description above,
there is a potential difficulty concerning the {\it physical}
realisation of an exponentially large unstructured database. One
might expect that it will require exponentially many degrees of
freedom of some physical resource, such as space, and consequently
it may need exponential (i.e. $O(N)$) effort or time just to access
a typical (remotely lying) record. We will replace the database by
an oracle which computes an $n$ bit function $f:B^n \rightarrow B$
(where $B=
\{ 0,1 \}$). It is promised that $f(x)=0$ for all $n$ bit strings
except exactly one string, denoted $x_0$ (the ``marked'' position)
for which $f(x_0 )=1$. Our problem is to determine $x_0$. We assume
as usual that $f$ is given as a unitary transformation $U_f$ on
$n+1$ qubits defined by \begin{equation} \label{uf} U_f
\ket{x}\ket{y} = \ket{x}\ket{y\oplus f(x)} \end{equation}
Here the input register $\ket{x}$ consists of $n$ qubits as $x$
ranges over all $n$ bit strings and the output register $\ket{y}$
consists of a single qubit with $y=0$ or 1. The symbol $\oplus$
denoted addition modulo 2.\\
\begin{picture}(14,4)(0,0.8)
\put(5,1){\framebox(2,3){$U_f$}}
\put(3,1.8){\vector(1,0){1}}
\put(4,1.8){\line(1,0){1}}
\put(3,3.2){\vector(1,0){1}}
\put(4,3.2){\line(1,0){1}}
\put(7,1.8){\vector(1,0){1}}
\put(8,1.8){\line(1,0){1}}
\put(7,3.2){\vector(1,0){1}}
\put(8,3.2){\line(1,0){1}}
\put(2,3.1){$\ket{x}$}
\put(2,1.7){$\ket{y}$}
\put(9.4,3.1){$\ket{x}$}
\put(9.4,1.7){$\ket{y \oplus f(x)}$}
\end{picture}
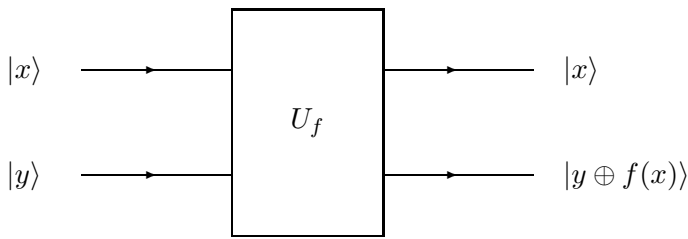
\\{\small Figure 2. The action of $U_f$ on a general basis state
$\ket{x}\ket{y}$ of the input and output registers.}\\[2mm] The
assumption that the database was unstructured is formalised here as
the standard oracle idealisation that we have no access to the
internal workings of $U_f$ -- it operates as a ``black box'' on the
input and output registers. In this formulation there is no problem
with the access to $f(x)$ for any of the exponentially many $x$
values and indeed we may also readily query the oracle with a
superposition of input values.

Instead of using $U_f$ we will generally use an equivalent
operation denoted $I_{x_0}$ on $n$ qubits. It is defined by
\begin{equation} \label{ixoo} I_{x_0}\ket{x} = \left\{ \begin{array}{rl}
\ket{x} & \mbox{if $x\neq x_0$ } \\ -\ket{x_0} & \mbox{if $x=x_0$}
\end{array} \right. \end{equation}
i.e. $I_{x_0}$ simply inverts the amplitude of the $\ket{x_0}$
component. If $x_0$ is the $n$ bit string $00\ldots 0$ then
$I_{x_0}$ will be written simply as $I_0$.

A black box which performs $I_{x_0}$ may be simply constructed from
$U_f$ by just setting the output register to $\frac{1}{\sqrt{2}} (
\ket{0}-\ket{1})$. Then the action of $U_f$ leaves the output
register in this state and effects $I_{x_0}$ on the input
register\footnote{Note that, conversely, $U_f$ may be constructed
from the a black box for $I_{x_0}$ as follows. Let $J_{x_0}$ denote
the operation ``$I_{x_0}$ controlled by the output qubit''. Apply
$H$ to the output register, then $J_{x_0}$ to the input and output
registers, then $H$ again to the output register. The total effect
is just $U_f$ on the $n+1$ qubits of both registers, as the reader
may verify. To construct $J_{x_0}$ from $I_{x_0}$ we also need an
eigenstate of $I_{x_0}$. The construction is described in
\cite{KIT} or \cite{JOZ98b}.}:\\
\begin{picture}(14,4)(0,0.8)
\put(5,1){\framebox(2,3){$U_f$}}
\put(3,1.8){\vector(1,0){1}}
\put(4,1.8){\line(1,0){1}}
\put(3,3.2){\vector(1,0){1}}
\put(4,3.2){\line(1,0){1}}
\put(7,1.8){\vector(1,0){1}}
\put(8,1.8){\line(1,0){1}}
\put(7,3.2){\vector(1,0){1}}
\put(8,3.2){\line(1,0){1}}
\put(2,3.1){$\ket{\psi}$}
\put(1.4,1.7){$\ket{0}-\ket{1}$}
\put(9.4,3.1){$I_{x_0}\ket{\psi}$}
\put(9.4,1.7){$\ket{0}-\ket{1}$}
\end{picture}
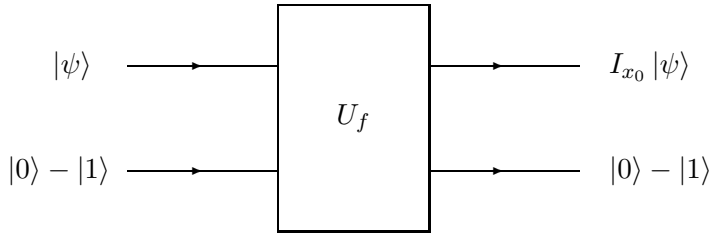
\\{\small Figure 3. Construction of $I_{x_0}$ from $U_f$. Here
$\ket{\psi}$ is any $n$-qubit state.}\\

Our searching problem becomes the following: we are given a black
box which computes $I_{x_0}$ for some $n$ bit string $x_0$ and we
want to determine the value of $x_0$.

\noindent
{\bf REFLECTIONS ON REFLECTIONS}

We first digress briefly to record some elementary properties of
reflections which will provide the basis for our interpretation of
Grover's algorithm. \\{\bf In two real dimensions:}\\In real 2
dimensional Euclidean space, let $M$ be any straight line through
the origin specified by a unit vector $v$ {\it perpendicular} to
$M$. Let $I_v$ denote the operation of reflection in $M$. Note that
if $u$ is any vector we may write it uniquely as a sum of
components parallel and perpendicular to $v$. If $v^\perp$ is a
unit vector lying along $M$ then we have
\[ u=av+bv^\perp \]
and $I_v$ simply replaces $a$ by $-a$.\\{\bf In $N$ complex
dimensions:}\\ Note that $I_v$ is exactly like $I_{x_0}$ in eq.
(\ref{ixoo}) above except that there, we were in a complex space of
higher dimension. We may interpret $I_{x_0}$ as a reflection in the
hyperplane orthogonal to $\ket{x_0}$. In terms of formulas, to
reflect the $x_0$ amplitude we have
\begin{equation} \label{ixo} I_{x_0}= I-2 \proj{x_0} \end{equation}
where $I$ is the identity operator. More generally if $\ket{\psi}$
is any state we define \begin{equation}\label{ipsi}
I_{\ket{\psi}}=I-2\proj{\psi} \end{equation} Then $I_{\ket{\psi}}$
is the operation of reflection in the hyperplane\footnote{More
generally, for any subspace $D$ we may define $I_D$ by $I_D =
I-2\sum_d \proj{d}$ where $\{\ket{d}\}$ is any orthonormal basis of
$D$. Then $I_D$ is reflection in the orthogonal complement
$D^\perp$ of the subspace $D$.} orthogonal to $\ket{\psi}$. For any
state $\ket{\chi}$ we may uniquely express it as a sum of
components parallel and orthogonal to $\ket{\psi}$ and
$I_{\ket{\psi}}$ simply inverts the parallel component.

We have the following simple properties of $I_{\ket{\psi}}$:\\{\bf
Lemma 1:} If $\ket{\chi}$ is any state then $I_{\ket{\psi}}$
preserves the 2-dimensional subspace ${\cal S}$ spanned by
$\ket{\chi}$ and $\ket{\psi}$.\\{\bf Proof:} Geometrically, $\cal
S$ and the mirror hyperplane are orthogonal to each other (in the
sense that the orthogonal complement of either subspace is
contained in the other subspace) so the reflection preserves $\cal
S$. Alternatively in terms of algebra, eq. (\ref{ipsi}) shows that
$I_{\ket{\psi}}$ takes $\ket{\psi}$ to $-\ket{\psi}$ and for any
$\ket{\chi}$, it adds a multiple of $\ket{\psi}$ to $\ket{\chi}$.
Hence any linear combination is mapped to a linear combination of
the same two states $\Box$.\\{\bf Lemma 2:} For any unitary
operator $U$
\[ UI_{\ket{\psi}} U^{-1} = I_{U\ket{\psi}} \] {\bf Proof:}
Geometrically we are just changing description (reference basis) by
$U^{-1}$ but the result is also immediate from eq. (\ref{ipsi}):
\[ UI_{\ket{\psi}} U^{-1} = I-2U\proj{\psi}U^{-1}
= I-2 \proj{U\psi} = I_{U\ket{\psi}} \hspace{1cm} \Box . \]
Looking back at eq. (\ref{qu}) we see that
\begin{equation} \label{qii} Q=-I_{U\ket{0}} I_{\ket{x_0 }}
\end{equation}
{\bf Back to two real dimensions:}\\ By lemma 1, both $I_{x_0}$ and
$I_{U\ket{0}}$ preserve the two dimensional subspace $\cal V$
spanned by $\ket{x_0}$ and $U\ket{0}$ . Hence by eq. (\ref{qii}),
$Q$ preserves $\cal V$ too. Now we may introduce a basis $\{
\ket{e_1}, \ket{e_2} \}$ into $\cal V$ such that $U\ket{0}$ and
$\ket{x_0}$ up to an overall phase, have {\it real} coordinates.
Indeed choose $\ket{e_1} = U\ket{0}$ so $U\ket{0}$ has coordinates
$(1,0)$. Then $e^{i\xi} \ket{x_0}= a\ket{e_1}+b\ket{e_2}$ where
$\ket{e_2}$, orthonormal to $\ket{e_1}$, still has an overall phase
freedom. Thus choose $\xi$ to make $a$ real and the phase of
$\ket{e_2}$ to make $b$ real. Then in this basis, since $U\ket{0}$
and $\ket{x_0}$ have real coordinates, the operators $I_{x_0}$ and
$I_{U\ket{0}}$ when acting on $\cal V$, are also described by {\it
real} 2 by 2 matrices -- in fact they are just the real 2
dimensional reflections in the lines perpendicular to $\ket{x_0}$
and $U\ket{0}$ in $\cal V$. Finally we have:\\{\bf Lemma 3:} For
any 2 dimensional real $v$ we have \[ -I_v = I_{v^\perp} \] where
$v^\perp$ is a unit vector perpendicular to $v$.\\{\bf Proof:} For
any vector $u$ we write $u=av+bv^\perp$. Then $I_v$ just reverses
the sign of $a$ and $-I_v$ reverses the sign of $b$. Thus the
action of $-I_v$ is the same as that of $I_{v^\perp}$ $\Box$.

Later, this lemma will explain the significance of the minus sign
in eq. (\ref{qu}). For the present, note that from eq. (\ref{qii})
we can write
\[ Q = I_{\ket{w}} I_{\ket{x_0}} \]
where $\ket{w}$ is orthogonal to $U\ket{0}$ and lies in the plane
of $U\ket{0}$ and $\ket{x_0}$. Since we are working with real
coordinates, theorem 1 shows that $Q$, acting in $\cal V$, is just
the operation of rotation through angle $2\alpha$ where $\alpha$ is
the angle between $\ket{w}$ and $\ket{x_0}$ i.e. $\cos \alpha =
\amp{x_0}{w}$. Since $U\ket{0}$ is perpendicular to $\ket{w}$ we
can write $\sin \alpha = \amp{x_0}{U|0}$.

\noindent
{\bf GROVER'S ALGORITHM}

We now give an interpretation of the workings of the quantum
searching algorithm in view of the preceeding simple facts about
reflections. Given the black box $I_{x_0}$ how can we identify
$x_0$? Surely we must apply $I_{x_0}$ to {\it some} state (we can
do nothing else with a black box!) but there seems no reason a
priori to choose any one state rather than any other. So let us
just choose a state $\ket{w}$ {\em at random}. $\ket{w}$ may be
written as $U\ket{0}$ where $U$ is chosen at random.

Now by lemma 1, $I_{x_0}$ preserves the subspace spanned by
$\ket{w}$ and $\ket{x_0}$ and by theorem 1, $I_{\ket{w}}I_{x_0}$
provides a way of moving around in this subspace -- it is just
rotation by twice the angle between $\ket{x_0}$ and $\ket{w}$.
(Note that $I_{\ket{w}}$ may be constructed via lemma 2 as $UI_0
U^{-1}$.) The idea now is to try to use this motion to move from
the known starting state $\ket{w}$ towards the unknown $\ket{x_0}$.
This process has been called ``amplitude amplification''
\cite{GRO97a,BRA98} as we are effectively trying to enhance the
amplitude of the $\ket{x_0}$ component of the state. Once we are
near to $\ket{x_0}$ then a measurement of the state in the standard
basis $\{ \ket{x}
\}$ will reveal the value of $x_0$ with high probability.

However there is an apparent problem: we do not know $x_0$ so we
know neither the angle $\beta$, between $\ket{x_0}$ and $\ket{w}$,
nor the angle $2\beta$ of rotation of $I_{\ket{w}}I_{x_0}$. Hence
we do not know how many times to apply the rotation to move
$\ket{w}$ near to $\ket{x_0}$. Remarkably we can solve this problem
by using the extra information that $\ket{x_0}$ is known to be a
member of a particular basis $\{ \ket{x} \}$ of $N$ orthonormal
states!  If we choose
\begin{equation} \label{w0} \ket{w_0}= \frac{1}{\sqrt{N}} \sum_x
\ket{x} \end{equation}
to be a {\it uniform} superposition of all the $\ket{x}$'s then
{\it whatever} the value of $x_0$ is, we have that $\amp{x_0}{w_0}=
\frac{1}{\sqrt{N}}$ and hence we will know that the angle $\beta$ is given
by $\cos \beta = \frac{1}{\sqrt{N}}$ in every possible case! Note
that for large $N$ (the usual case of interest) $\ket{x_0}$ and
$\ket{w_0}$ are nearly orthogonal so $2\beta$ is near to $\pi$.
This will typically be the case for any $\ket{w}$ chosen at random
in a large Hilbert space -- it will tend with high probability to
be nearly orthogonal to any previously fixed state such as
$\ket{x_0}$.

Now, $I_{\ket{w}}I_{x_0}$ rotating through nearly $\pi$, acts
rather wildly on the space, moving vectors through a great distance
and we would prefer to have a gentler incremental motion of
$\ket{w_0}$ towards $\ket{x_0}$. One way of doing this is to use
instead, the operation $(I_{\ket{w_0}}I_{x_0})^2$ rotating through
$4\beta$ which is near to $2\pi$ i.e. 0 mod $2\pi$. Since $\cos
\beta =
\frac{1}{\sqrt{N}}$
we have that $4\beta$ mod $2\pi$ is $O(\frac{1}{\sqrt{N}})$. (To
see this, write $\beta = \frac{\pi}{2}-\alpha$ so $\sin \alpha
= \cos \beta = \frac{1}{\sqrt{N}}$. Then $4\beta = 2\pi -4\alpha$
and $\alpha \approx \frac{1}{\sqrt{N}}$ for large $N$). Now the
angle between $\ket{w_0}$ and $\ket{x_0}$ is nearly $\pi
/2$ so we will need $O(\sqrt{N})$ iterations of this rotation to move
$\ket{w_0}$ near to $\ket{x_0}$. A second way of dealing with the
large $\beta$ problem -- the way actually used in Grover's
algorithm -- is to simply put a {\it minus} sign in front of
$I_{\ket{w_0}}I_{x_0}$! This explains the role of the minus sign in
eq. (\ref{qu}). Indeed by lemma 3,
$-I_{\ket{w}}I_{x_0}=I_{\ket{w_0^\perp}}I_{x_0}$ where
$\ket{w_0^\perp}$ is orthogonal to $\ket{w_0}$ in the subspace
spanned by $\ket{w_0}$ and $\ket{x_0}$ and now the angle $\alpha$
between $\ket{w_0^\perp}$ and $\ket{x_0}$ is given by $\cos
\alpha = \amp{x_0}{w^\perp}$ i.e. $\sin \alpha = \amp{x_0}{w}$ so
$\alpha \approx \frac{1}{\sqrt{N}}$. Again we will need
$O(\sqrt{N})$ iterations of the rotation $-I_{w_0}I_{x_0}$ through
$2\alpha$ to span the angle between $\ket{w_0}$ and $\ket{x_0}$.

In conclusion, we choose the starting state $\ket{w_0}$ of eq.
(\ref{w0}) and apply $O(\sqrt{N})$ times, the operator
\[ Q = I_{\ket{w_0^\perp}}I_{x_o} = - I_{\ket{w_0}} I_{x_0}
= - UI_0 U^{-1} I_{x_0} \]
where $U$ is any unitary operation with $U\ket{0}=\ket{w_0}$ (for
example $U=H$). The significance of the minus sign (in the context
of reflection operations) is to convert a nearly orthogonal pair of
directions to a nearly parallel pair (c.f. lemma 3) . The composite
structure of $Q$ is just to build a rotation as a product of two
reflections (c.f. theorem 1) and the random choice of $U$ just
picks a random starting state in the two dimensional subspace,
which is then moved towards $\ket{x_0}$. The exact number of
iterations of $Q$ required depends on knowledge of the angle
between $\ket{x_0}$ and $U\ket{0}$. If $U\ket{0}=
\ket{w_0}=\frac{1}{\sqrt{N}} \sum_x
\ket{x}$ then this is explicitly known, but if $U$ is chosen more
generally at random then we will not know this angle. However $Q$
will still generally be a small rotation through some angle of
order $O(\frac{1}{\sqrt{N}})$ but we will not know when to stop the
iterations. Nevertheless the process will still move $U\ket{0}$
near to $\ket{x_0}$ in $O(\sqrt{N})$ steps. It will fail only in
the unlikely situation that the randomly chosen $U\ket{0}$ happens
to be {\it exactly} orthogonal to the unknown $\ket{x_0}$ (i.e. $U$
has zero matrix element $U_{0x_0}$ as noted in
\cite{GRO97a,GRO98}). $Q$ is then a rotation through an angle of
zero.

Having identified $W$ as a rotation through $2\alpha$ in the plane
$\cal V$  of $\ket{x_0}$ and $\ket{w_0}$, where $\alpha$ is defined
by
\begin{equation} \label{alp} \sin
\alpha = \frac{1}{\sqrt{N}}\end{equation}
(for the case that $U\ket{0}=\ket{w_0}$) we may readily calculate
the motion of $\ket{\psi_0}
=
\ket{w_0}=\frac{1}{\sqrt{N}}\sum_x \ket{x}$ towards $\ket{x_0}$ by
iterated application of $W$. In $\cal V$ introduce the basis $\{
\ket{x_0}, \ket{x_0^\perp } \}$ where $\ket{x_0^\perp}
=\frac{1}{\sqrt{N-1}} \sum_{x\neq x_0} \ket{x}$. Then from eq.
(\ref{alp}) we get \[ \ket{\psi_0} = \sin \alpha \ket{x_0}+\cos
\alpha \ket{x_0^\perp} \] If we write \[ \ket{\psi_{n+1}} = Q\ket{\psi_n}
\hspace{3mm} \mbox{and} \hspace{3mm}
\ket{\psi_n}= \sin \alpha_n \ket{x_0}+\cos \alpha_n
\ket{x_0^\perp} \] then we immediately get from the interpretation of $Q$
as a rotation through $2\alpha$
\[ \alpha_{n+1}=\alpha_n + 2\alpha \hspace{1cm} \mbox{i.e.\,\,\, $
\alpha_n = (2n+1)\alpha$} \]
giving the solution of the iteration derived in \cite{BOY96}. For a
more general choice $\ket{\psi_0}=\ket{w}$ of starting state, the
value of $\alpha$, given by $\sin^{-1} |\amp{x_0}{w}|$, will
generally be unknown but the application of $W$ still increments
the angle successively by $2\alpha$ as above. The number of
iterations is chosen to make $\alpha_n$ as close as possible to
$\pi /2$.

\noindent
{\Large\bf Acknowledgement}

\noindent
After this note was completed it came to my attention that various
other workers were aware of the significance of double reflections
for Grover's algorithm. However it appears not to be widely known
and I present this note for its pedagogical value.

\end{document}